\documentclass[prb,superscriptaddress,twocolumn,showpacs,amsmath,amssymb,floatfix,groupedaddress]{revtex4-1}

\usepackage{graphicx}
\usepackage{xspace}
\newcommand{\dto}{$\text{Dy}{}_\text{2}\text{Ti}_\text{2}\text{O}_\text{7}$\xspace}
\newcommand{\yto}{$\text{Y}{}_\text{2}\text{Ti}_\text{2}\text{O}_\text{7}$\xspace}
\newcommand{\ruo}{$\text{Ru}\text{O}_\text{2}$\xspace}
\newcommand{\zz}{\mbox{2in-2out}\xspace}
\newcommand{\ed}{\mbox{1in-3out}\xspace}
\newcommand{\de}{\mbox{3in-1out}\xspace}
\newcommand{\beee}{\mbox{$\vec B \,||\, [111]$}\xspace}
\newcommand{\bnne}{\mbox{$\vec B \,||\, [001]$}\xspace}
\newcommand{\been}{\mbox{$\vec B \,||\, [110]$}\xspace}
\newcommand{\kagome}{Kagom\'e\xspace}

\newcommand{\kapparel}{\mbox{$\kappa/\kappa_0$}\xspace}
\newcommand{\kapparellong}{\mbox{$\kappa(B)/\kappa(0\,\text{T})$}\xspace}
\newcommand{\kappamag}{\mbox{$\kappa_\text{mag}$}\xspace}
\newcommand{\kappaph}{\mbox{$\kappa_\text{ph}$}\xspace}
\newcommand{\kappaphb}{\mbox{$\kappa_\text{ph}(B)$}\xspace}

\newcommand{\simm}[2]{\mbox{$\sim #1\,\text{#2}$}\xspace}
\newcommand{\dytolong}{\mbox{$(\text{Dy}{}_\text{0.5}\text{Y}{}_\text{0.5})_\text{2}\text{Ti}_\text{2}\text{O}_\text{7}$}\xspace}
\newcommand{\cf}{\emph{cf.}\xspace}
\newcommand{\ie}{\emph{i.e.}\xspace}
\newcommand{\eg}{\emph{e.g.}\xspace}

\begin{document}

\title{Anisotropic Monopole Heat Transport in the Spin-Ice Compound \dto}

\author{G.~Kolland}
\author{M.~Valldor}
\author{M.~Hiertz}
\author{J.~Frielingsdorf}
\author{T.~Lorenz}\email[E-mail: ]{tl@ph2.uni-koeln.de}

\affiliation{{\protect II.}\ Physikalisches Institut, Universit\"at zu K\"oln, Z\"ulpicher Str.\ 77, 50937 K\"oln, Germany}

\begin{abstract}
  We report a study of the thermal conductivity $\kappa$ of the spin-ice material \dto.
  From the anisotropic magnetic-field dependence of $\kappa$ and by additional measurements on the phononic reference compounds \yto and \dytolong,
  we are able to separate the phononic and the magnetic contributions to the total heat transport, \ie \kappaph and \kappamag, respectively,
  which both depend on the magnetic field.
  The field dependent \kappaphb arises from lattice distortions due to magnetic-field induced torques on the non-collinear magnetic moments of the Dy ions.
  For \kappamag, we observe a highly anisotropic magnetic-field dependence, which correlates with the corresponding magnetization data
  reflecting the different magnetic-field induced spin-ice ground states.
  The magnitude of \kappamag increases with the degree of the ground-state degeneracy.
  This anisotropic field dependence as well as various hysteresis effects suggest
  that \kappamag is essentially determined by the mobility of the magnetic monopole excitations in spin ice.         
\end{abstract}

\pacs{
66.70.-f, 
75.40.Gb 
75.47.-m 
}

\date{\today}

\maketitle

\section{INTRODUCTION}

  \dto is a geometrically frustrated spin system with a degenerate, so-called spin-ice ground state.
  The magnetic Dy sites in \dto form a pyrochlore lattice, consisting of corner-sharing tetrahedra.
  A strong crystal field results in an Ising anisotropy of the magnetic moments of the Dy ions,
  which align along their local easy axes in one of the \{111\}~directions
  and point either into or out of the tetrahedra.
  Possible ground states in zero magnetic field are given by the ``ice~rule'': two spins point into and two out of a tetrahedron.
  This behavior is analogous to the hydrogen displacement in water ice revealing a residual zero-temperature entropy \cite{Ramirez1999,Bramwell2001,Hiroi2003,Sakakibara2003,Nagle1966}.
  Magnetic excitations can be created pairwise by flipping one spin resulting in two neighboring tetrahedra with configurations \ed and \de, respectively.
  Such a dipole excitation can fractionalize into two individual monopole excitations which can freely propagate in zero magnetic field \cite{Ryzhkin2005,Castelnovo2008,Morris2009,Giblin2011,Castelnovo2011,PhysRevLett.105.267205,Kadowaki2009,Jaubert2011,Bramwell2009,Blundell2012,Matsuhira2011,Yaraskavitch2012}.

  Recently, we showed that these exotic magnetic excitations contribute to the heat transport in \dto \cite{Kolland2012}.
  At temperatures around \simm{0.6}{K}, the magnetic contribution \kappamag accounts for almost 50\% of the total thermal conductivity~$\kappa$.
  A magnetic field \bnne lifts the ground-state degeneracy of the magnetic system in \dto.
  This results in a complete suppression of \kappamag for rather small magnetic fields of \simm{0.5}{T}.
  An alternative interpretation had been proposed earlier in Ref.~\onlinecite{Klemke2011}, where the thermal conductivity was assumed to be
  purely phononic and the field dependence of $\kappa$ was attributed to phonon scattering on magnetic excitations.
  Our data, however, do not support this interpretation (\cf~discussion in Ref.~\onlinecite{Kolland2012}).
  In Ref.~\onlinecite{Kolland2012}, we studied \kappamag for a magnetic field applied parallel to $[001]$.
  By applying the magnetic field parallel to $[111]$ or $[110]$, one can realize more complex field-induced ground states with different degrees of degeneracy.
  For \beee, the field-induced ground state below \mbox{1 T} (\kagome-ice state)
  is 3-fold degenerate consisting of different types of \zz tetrahedra \cite{Tabata2006,Matsuhira2002}.
  Above \mbox{1 T}, the ground state changes into a non-degenerate ground state consisting of alternating \de and \ed tetrahedra.
  For a magnetic field parallel to $[110]$, two spins per tetrahedron are perpendicular to $\vec B$ and are, thus, not affected.
  This leads to a 2-fold degenerate ground state, even for arbitrarily large magnetic fields.
  In this paper, we study the dependence of the magnetic heat transport on the degree of the field-induced ground-state degeneracy
  realized by the field directions $[001]$, $[111]$, and $[110]$.

  In the course of this study, we also extend the field-dependent thermal-conductivity measurements of \dto to magnetic fields up to \mbox{7 T}.
  An additional field dependence of $\kappa$ is observed for higher fields (depending on the actual field direction)
  which we identify as a field-dependent phononic background \kappaphb of \dto.
  The phononic background is essentially given by the field-dependent $\kappa(B)$ of the Y-doped \dytolong.
  This compound can be regarded as a magnetic reference system with strongly suppressed spin-ice features.
  Compared to the non-magnetic \yto, the half-doped reference compound can be utilized to study the magnetic-field dependence
  of the phononic contribution \kappaph of \dto.
  The field dependence of \kappaph can be explained by a magnetostrictive lattice distortion originating from
  torques of the Dy~momenta induced by the external magnetic field.

\section{EXPERIMENTAL}

  Single crystals were grown from sintered bars of \mbox{$\text{TiO}_2$} (3N, Sigma-Aldrich) and \mbox{$\text{Y}_2\text{O}_3$}
  (4N, Alfa Aesar) and/or \mbox{$\text{Dy}_2\text{O}_3$} (4N, REacton) in proper stoichiometry.
  The sintering processes were done in air at \mbox{$1400\,^\circ\text C$} over night using corundum boats as crucibles.
  For the crystal growth, a floating-zone technique was applied inside a four-mirror image furnace
  to acquire centimeter-sized single crystals.
  To obtain crystals without cracks, the floating zone was run twice trough the bars:
  first at a rate of \mbox{$25\,\text{mm}/\text h$} and subsequently with \mbox{$7\,\text{mm}/\text h$}.
  Ambient pressure of pure oxygen was chosen as atmosphere during the growth.
  The pale yellow crystals exhibit high reflectance at a grazing angle but otherwise full transparency.
  \dto\ is slightly more yellow than \dytolong and \yto.
  The resulting crystals were found to be single domain by extensive Laue-photo investigations and the bulk purity was checked by
  X-ray powder diffraction using a STOE D5000 diffractometer, \mbox{Cu $\text K_\alpha$} radiation, and Bragg-Brentano reflection mode.
  The correct composition was confirmed by energy-dispersive spectroscopy inside a scanning electron microscope.

  The thermal conductivity was measured using the standard steady-state method.
  The temperature difference was produced by a heater attached at one end of the sample
  and measured by a pair of matched \ruo~thermometers.
  The thermal-conductivity measurements were performed on bar-shaped single crystals of approximately \mbox{$3\times1\times1\,\text{mm}^3$}
  with the long edge parallel to the
  $[1\bar10]$~direction. The heat current was directed along the long edge of the crystal.
  In most cases, the magnetic fields applied along the different directions were oriented perpendicular to this long edge.
  Hence, considerable demagnetization effects have to be taken into account.
  The demagnetization field is calculated on the basis of experimental magnetization data
  which were measured with a home-built Faraday magnetometer on thin samples to minimize demagnetization effects within the magnetization measurements.
  The magnetostriction measurements were done with a home-built capacitance dilatometer, which allows to measure the uniaxial length changes either parallel or perpendicular to the applied magnetic field.

\section{RESULTS}

  \begin{figure}
    \includegraphics[width=\linewidth]{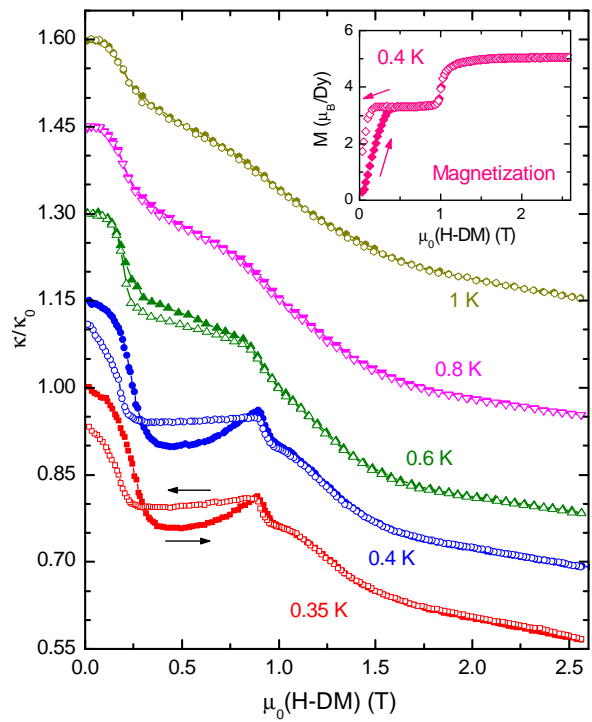}
    \caption{\label{kvb_b111}(Color online) Relative change \kapparel of \dto\ for \beee at various constant temperatures.
    The different curves are shifted by 0.15 with respect to each other.
    Inset: Magnetization $M(B)$ of \dto at \mbox{0.4 K}.
    The arrows indicate the field-sweep direction.
    Demagnetization effects are taken into account.}
  \end{figure}

  Fig.~\ref{kvb_b111} shows the relative change of the thermal conductivity \kapparellong of \dto for \beee at temperatures between \mbox{0.35 K} and \mbox{1 K}.
  The different datasets are shifted with respect to each other.
  With increasing field, $\kappa(B)$ decreases and exhibits a plateau within the \kagome-ice phase below \mbox{1 T}.
  For higher fields, $\kappa(B)$ further decreases and shows a kink around \mbox{1.5 T}.
  The plateau feature is more pronounced at lowest temperature and broadens towards higher temperature.
  The inset of Fig.~\ref{kvb_b111} displays the magnetization $M(B)$ of \dto for \beee at \mbox{0.4 K},
  which is in agreement with previous studies, \eg~Refs.~\onlinecite{Matsuhira2002,Sakakibara2003}.
  The magnetization as well as the thermal conductivity have been measured with slow sweep rates \cite{footnote2}
  to avoid thermal-runaway effects, \cf~Ref.~\onlinecite{PhysRevLett.105.267205}.
  The basic features of the $\kappa(B)$~data are reflected within the magnetization data,
  which also exhibit a pronounced plateau within the \kagome-ice phase.
  At \mbox{1 T}, $M(B)$ shows a sharp kink and increases towards the saturation value, which is essentially reached at \simm{1.5}{T}.
  This increase of $M(B)$ is accompanied by a decrease of $\kappa$ between \mbox{1 T} and \mbox{1.5 T}.
  The hysteresis in $M(B)$ at low temperature (\simm{0.4}{K}) below \simm{0.3}{T} can also be observed in the thermal-conductivity data.
  Even the remnant zero-field magnetization observed in the measurement with decreasing field (open symbols),
  which results from slow relaxation processes of the spin ice, is reflected within the thermal-conductivity data.
  Below \simm{0.4}{K}, the $\kappa$~data for decreasing field end in a reduced zero-field value,
  which slowly relaxes back to the initial zero-field value (after zero-field cooling).
  Such a correlation of magnetization and thermal conductivity has already been observed for \bnne (Ref.~\onlinecite{Kolland2012}).
  For \beee, however, clear differences between $M(B)$ and $ \kappa(B)$ are also found.
  First of all, at lowest temperature, $\kappa(B)$ within the \kagome-ice phase strongly depends on whether the measurement is performed
  with increasing field (after cooling in zero field) or with decreasing field (starting from high magnetic fields).
  Such a hysteresis is not present in the magnetization data.
  Even at lowest temperature, the magnetization within the \kagome-ice phase shows no hysteresis with respect to the field-sweep direction.
  Secondly, $M(B)$ and $\kappa(B)$ show different high-field behaviors.
  At \mbox{0.4 K}, the magnetization is almost saturated above \mbox{1.5 T}, whereas the thermal conductivity $\kappa(B)$ further decreases with increasing field even above \mbox{1.5 T}.
  The question arises whether these effects can be attributed to the phononic or to the magnetic contribution of $\kappa$.
  
  \begin{figure}
    \includegraphics[width=\linewidth]{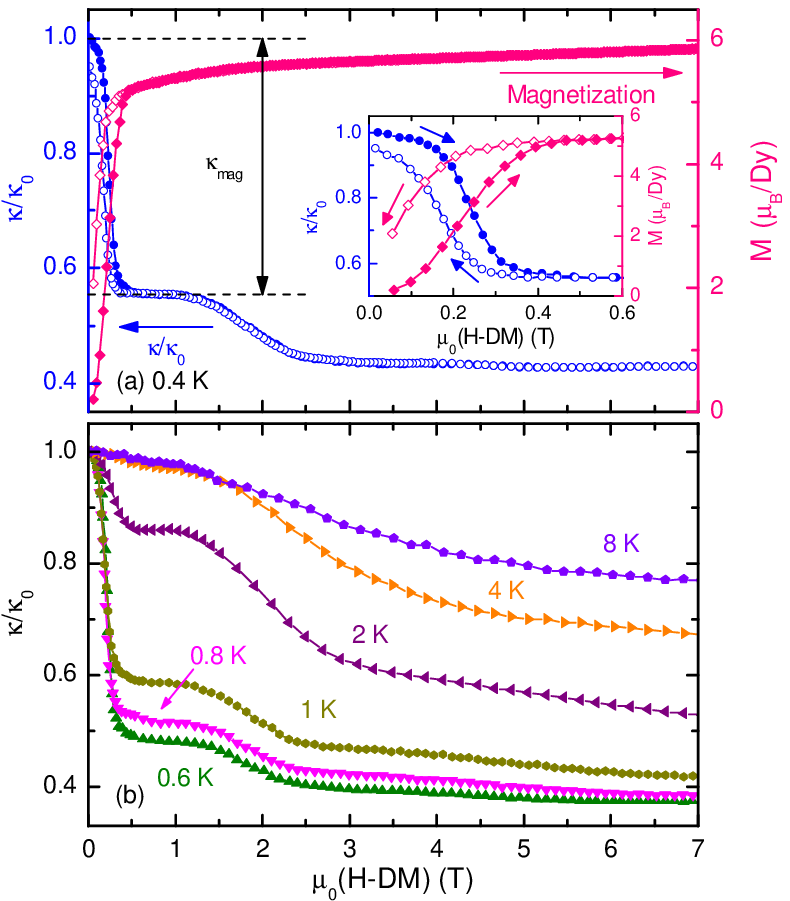}
    \caption{\label{kvb_b001}(Color online) Relative change \kapparel of \dto for \bnne at various constant temperatures between \mbox{0.4 K} and \mbox{8 K}.
    Demagnetization effects are taken into account.
    Panel (a) compares the field dependencies of magnetization and thermal conductivity at \mbox{0.4 K}. The complete
    suppression of \kappamag below 1~T is tagged by the dashed lines. The 
    inset is an enlargement of the low-field region, where the field-sweep directions are marked by arrows.}
  \end{figure}

  In order to analyze this, we will first discuss the high-field dependence of $\kappa$ for magnetic fields above \simm{1.5}{T}.
  Because the magnetization is essentially saturated above \mbox{1.5 T},
  the magnetic system in \dto is fully polarized.
  Hence, one would expect the field dependence of $\kappa$ above \mbox{1.5 T} to
  originate from a change of \kappaph rather than from \kappamag.
  By means of the data shown in Fig.~\ref{kvb_b111}, however, one cannot straightforwardly separate \kappamag and \kappaph.
  In Ref.~\onlinecite{Kolland2012}, we showed that \kappamag vanishes for a rather small field of \simm{0.5}{T} applied parallel to $[001]$.
  Compared to \beee, the field-induced ground state for \bnne is less complex, as the ground-state degeneracy is lifted in
  one single step.
  Therefore, we start with the analysis of the high-field thermal conductivity of \dto for \bnne.
  The data published in Ref.~\onlinecite{Kolland2012} were only measured up to
  a (demagnetization-corrected) maximum field of \mbox{0.5 T}.
  Here, we extend the present $\kappa$~data to magnetic fields up to \mbox{7 T}.
  The results are displayed in Fig.~\ref{kvb_b001}, where the relative change \kapparellong is shown for temperatures
  between \mbox{0.4 K} and \mbox{8 K}.
  Below \mbox{0.5 T}, the data shown in Fig.~\ref{kvb_b001} confirm the data previously published in Ref.~\onlinecite{Kolland2012}.
  Fig.~\ref{kvb_b001}(a) directly compares the field-induced change of $\kappa$ with the magnetization $M(B)$
  at the lowest measured temperature of \mbox{0.4 K}.
  Below \mbox{0.5 T}, $\kappa(B)$ steplikely decreases.
  As can be seen in the inset of panel~(a), the sharp drop of $\kappa(B)$ below \mbox{0.5 T}
  clearly correlates with $M(B)$ and can, thus, be attributed to \kappamag.
  Above \simm{1.5}{T}, $\kappa(B)$ further decreases.
  This additional field dependence $\kappa(B)$, however, cannot be attributed to \kappamag as all spins are fully polarized.
  This is shown exemplarily for \mbox{0.4 K}, where the magnetization is essentially saturated above \mbox{1.5 T}
  (apart from a slight linear increase, see below).
  The field dependence of $\kappa$ also cannot be explained by phonon scattering on magnetic excitations as the excitation gap monotonically increases with
  an increasing field \bnne.
  Fig.~\ref{kvb_b001}(b) shows the field-induced relative change of $\kappa(B)$ at higher temperatures up to \mbox{8 K}.
  Above \simm{4}{K}, \kappamag vanishes, whereas a distinct high-field dependence of $\kappa$ above \mbox{1 T}
  is still observed even at \mbox{8 K}.
  As will be shown below, this high-field dependence of the thermal conductivity of \dto can be attributed to a magnetic-field dependent phononic contribution \kappaphb.
  
  \begin{figure}
    \includegraphics[width=\linewidth]{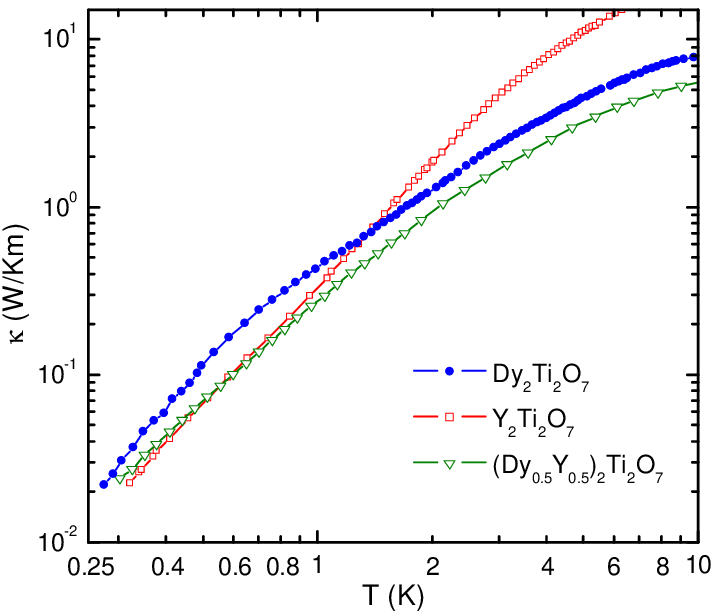}
    \caption{\label{kvt_dyto}(Color online) Comparison of the zero-field thermal conductivity of \dto (closed blue circles)
    with the non-magnetic reference \yto (open red squares)
    and the half-doped \dytolong (open green triangles).}
  \end{figure}

  In our previous work \cite{Kolland2012}, we studied the non-magnetic \yto as a purely phononic reference compound
  to the magnetic spin ice \dto.
  At temperatures below \simm{3}{K}, the thermal conductivity of \yto shows a very similar power-law behavior as the phononic background of
  \dto which was obtained by applying a magnetic field of \mbox{0.5 T} parallel to [001].
  For higher temperatures, however, the thermal conductivity of \dto is significantly smaller (see Fig.~\ref{kvt_dyto}).
  Most likely, this suppression originates from additional phonon scattering on the crystal-field excitations of the 4f~electrons of the Dy~ions.
  Here, we present an alternative approach in order to study the phonon background of $\kappa$ by using the half-doped compound \dytolong. 
  As every second Dy~ion is replaced by a non-magnetic Y~ion, {\it i.e.} on average two of the four Dy~ions per tetrahedron are replaced by Y~ions, the spin-ice properties are supposed to be strongly suppressed.
  However, due to the similar ionic radii of \mbox{$\text{Dy}^{3+}$} and \mbox{$\text{Y}^{3+}$}, one can assume both compounds to have comparable phononic properties.
  Fig.~\ref{kvt_dyto} shows the zero-field thermal conductivity of the spin-ice compound \dto together with the
  thermal conductivity of the non-magnetic reference \yto and the half-doped \dytolong.
  In all measurements, the heat current has been directed along the $[1\bar10]$~direction.
  Below \mbox{1 K}, both phononic reference compounds have very similar $\kappa$ values well below $\kappa$ of \dto.
  At higher temperature, the half-doped compound shows a behavior similar to the spin-ice compound, rather than to the non-magnetic
  \yto, which has significantly larger $\kappa$~values.
  This supports our interpretation of an additional phonon scattering on the crystal-field excitations of the Dy~ions, as given above.
  This additional scattering, indeed, can also be observed for the half-doped \dytolong, leading to $\kappa$~values close to the values of \dto.
  The slightly smaller values of the half-doped compound can be explained by an enhanced defect scattering of phonons
  which originates from the partial Dy~substitution.

  \begin{figure}
    \includegraphics[width=\linewidth]{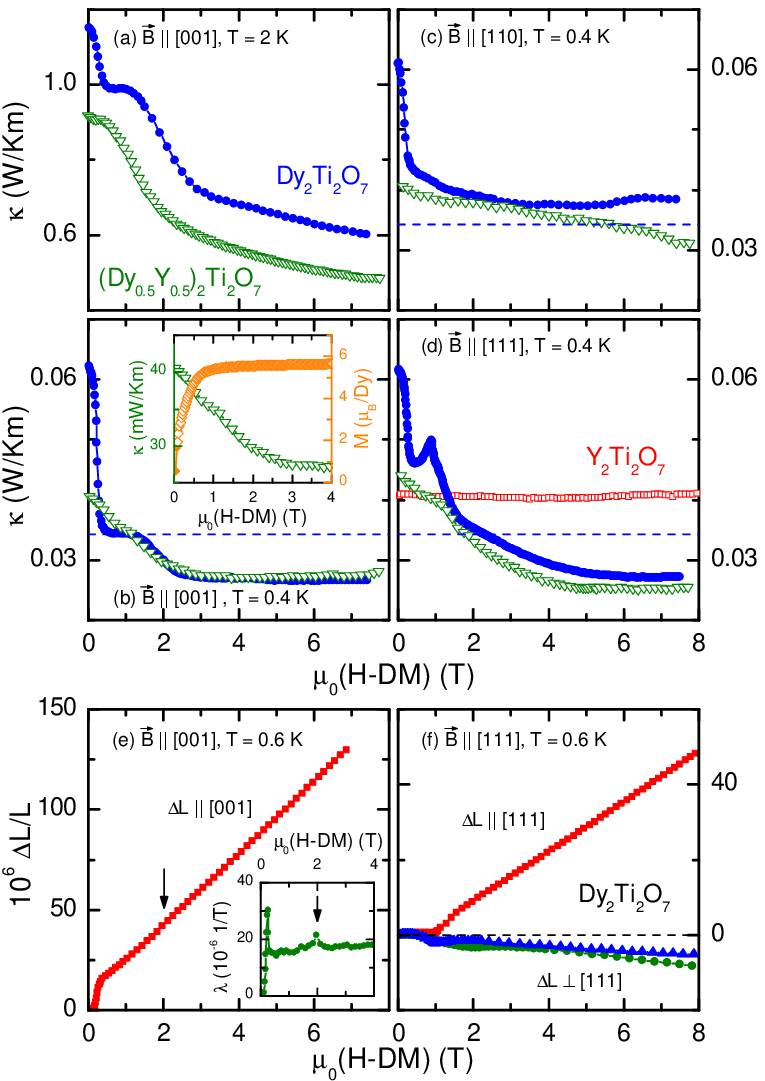}
    \caption{\label{kvb_dyto}(Color online) (a)-(d): Field-dependent $\kappa(B)$ of \dto (blue circles) and of the half-doped phononic reference compound \dytolong
    (green triangles) for different magnetic-field directions; panel (d) also contains $\kappa(B)$ of \yto .
    The inset in panel (b) compares $\kappa(B)$ with $M(B)$ for \dytolong.
    (e) and (f): Field-induced length changes of \dto for \bnne and \beee, respectively.
    The inset in panel (e) shows the corresponding field derivative of $\Delta L/L$, where the anomaly 
    at 2~T (arrows) is better seen. In all cases, demagnetization effects are taken into account and, for clarity, only data  
    obtained with increasing magnetic field are shown.}
  \end{figure}

\section{DISCUSSION}  
  
\subsection{Phononic Heat Conductivity $\boldsymbol\kappa_\text{ph}$}  

  The main advantage of the half-doped compound \dytolong, compared to \yto, is the fact that
  it is still strongly magnetic and can, therefore, be utilized to study the magnetic-field dependence of the phononic background \kappaphb of \dto.
  As shown exemplarily for \beee in Fig.~\ref{kvb_dyto}(d), $\kappa$ of \yto (red squares) has no magnetic-field dependence, as expected for a non-magnetic compound.
  Figs.~\ref{kvb_dyto}(a)-(d) compare the thermal conductivity $\kappa(B)$ of \dto (blue circles) and of \dytolong (green triangles) for various magnetic-field directions,
  where the heat current $\vec j$ is always directed parallel to $[1\bar10]$.
  First of all, even at lowest temperature, $\kappa(B)$ of \dytolong is not hysteretic with respect
  to the field-sweep direction (not shown).
  This is contrary to the undoped \dto,
  where at lowest temperature (\simm{0.4}{K}),
  a clear hysteresis in $\kappa(B)$ is observed below \simm{0.3}{T} for all considered field directions,
  see \eg~Figs.~\ref{kvb_b111} and \ref{kappamag}.
  
  The data in Figs.~\ref{kvb_dyto}(a) and (b) are measured with a magnetic field parallel to [001].
  At \mbox{2 K} [Fig.~\ref{kvb_dyto}(a)], $\kappa(B)$ of \dto shows two steplike anomalies at \simm{0.5}{T} and \simm{2}{T}
  and continuously decreases above the second anomaly.
  The step at \simm{2}{T} and the subsequent continuous decrease are very well reflected by the $\kappa(B)$~data of \dytolong.
  The first anomaly (at \simm{0.5}{T}), which we attribute to the suppression of \kappamag in \dto, is, however, absent in the \dytolong~data.
  Fig.~\ref{kvb_dyto}(b) shows the data for the same field direction \bnne at a lower temperature of \mbox{0.4 K}.
  Here again, the high-field behavior is very similar for both compounds,
  whereas the sharp decrease below \simm{0.5}{T} is absent in the \dytolong~data.
  The inset of Fig.~\ref{kvb_dyto}(b) illustrates the different origins of the field dependencies of the half-doped compared to the undoped compound.
  One can very clearly see that for \dytolong, the change of $\kappa(B)$ does not correlate with the magnetization $M(B)$, which is essentially saturated at \simm{1}{T},
  whereas $\kappa(B)$ shows a distinct field dependence up to \simm{3}{T}.
  This is different from the case of \dto, where the change of $M(B)$ at low field (below \simm{0.5}{T})
  directly correlates with the suppression of \kappamag [inset of Fig.~\ref{kvb_b001}(a)].
  Thus, we conclude that the high-field decrease of $\kappa$ of \dto can be identified as a field-dependent phononic background \kappaphb.
  
  It is notable, however, that the zero-field $\kappa$~values of \dytolong at lowest temperature are larger than the
  phononic background of \dto obtained from the plateau value around \mbox{1 T} [Fig.~\ref{kvb_dyto}(b)].
  Most likely, this originates from a small remnant magnetic contribution \kappamag in \dytolong, which is more pronounced
  at lower temperature (\mbox{0.4 K}) than at higher temperature [\mbox{2 K}, Fig.~\ref{kvb_dyto}(a)].
  Hence, one cannot directly identify the low-field (below \simm{0.5}{T}) $\kappa$~values of the half-doped compound
  as the phononic background \kappaph of the undoped mother compound.
  Nevertheless, the remnant magnetic contribution in \dytolong is strongly suppressed compared to \kappamag in \dto,
  as expected for half doping.

  In Fig.~\ref{kvb_dyto}(c), the field dependence of $\kappa$ is shown for a magnetic field parallel to $[110]$.
  For this particular field direction, two of the four spins per tetrahedron in \dto are perpendicular to $\vec B$ and are, thus, not affected.
  This results in a 2-fold degenerate ground state for this field direction and we find that the decline of $\kappa(B)$ below \mbox{0.5 T} is less pronounced as compared to \bnne [dashed line, \cf~panel~(b)].
  This indicates that \kappamag cannot be completely suppressed by a field applied parallel to $[110]$.
  Furthermore, above \simm{0.5}{T}, $\kappa$~is hardly field dependent.
  This is different from the other considered field directions, where $\kappa(B)$ shows a distinct high-field dependence of $\kappa$.
  However, here also, the thermal conductivity of the half-doped reference compound closely resembles the high-field data of \dto
  at least up to \mbox{4 T}, whereas there are some deviations appearing at higher field.
  The data in Fig.~\ref{kvb_dyto}(c) might suggest that the zero-field $\kappa$~value of \dytolong
  essentially reflects the phononic background for \dto.
  This is, however, misleading as both compounds exhibit a non-vanishing zero-field magnetic contribution to $\kappa$ (as shown above for \bnne) and, obviously, the zero-field $\kappa$ is independent on the magnetic-field direction.
  Consequently,
  the field-induced 2-fold degenerate ground state realized by \been
  leads to a considerable magnetic contribution \kappamag, which is hardly field dependent up to highest fields of \mbox{7 T}.
  
  More complex ground states are realized by a magnetic field applied parallel to $[111]$.
  Within the \kagome-ice phase below \mbox{1 T}, the ground state is 3-fold degenerate.
  Above the transition at \mbox{1 T}, the magnetic system changes to a non-degenerate, fully polarized ground state.
  The highly degenerate \kagome-ice state results in a pronounced plateau within the field-dependent change of $\kappa$
  observed for \dto at lowest temperature (Fig.~\ref{kvb_b111}).
  The high-field dependence of $\kappa$ for \beee is shown in Fig.~\ref{kvb_dyto}(d) together with the field-dependent data
  for the half-doped \dytolong and the non-magnetic \yto.
  As mentioned above, \yto can only be used to estimate the zero-field phononic contribution,
  as $\kappa$ of \yto shows no field dependence.
  In complete analogy to the field directions discussed above, the field dependence of $\kappa$ above \simm{2}{T} is very well described
  by $\kappa$ of the half-doped reference compound, whereas the spin-ice features at lower field are almost completely suppressed.
  This, again, confirms our interpretation of $\kappa(B)$ of the half-doped reference compound as the field-dependent phononic background of \dto.
  However, it is more difficult to distinguish between the magnetic and the phononic contributions of \dto for \beee.
  The field dependence well above \simm{1.5}{T} can certainly be attributed to \kappaph, but 
  around \mbox{1.5 T}, \kappaphb and the suppression of \kappamag due to the lifting of the ground-state degeneracy above the 
  \kagome-ice phase overlap.
  Consequently, the suppression of \kappamag appears as a kink in $\kappa(B)$ for \beee, rather than a pronounced plateau, as observed for \bnne.
  The kink is located close to the zero-field phononic background obtained from the \bnne~data
  [horizontal dashed line, \cf~Fig.~\ref{kvb_dyto}(b)].
  This suggests that the field dependence \kappaphb below \simm{1.5}{T} 
  is much smaller than the suppression of \kappamag.
  Hence, when we restrict the discussion to the low-field region below \mbox{1.5 T}, we can assume a
  constant, \ie field-independent, phononic background \kappaph of \dto.
  The magnetic contributions \kappamag extracted on the basis of this assumption are discussed in the following subsection.

  A possible explanation of the field dependence of the phononic background \kappaph of \dto
  is given in terms of field-induced lattice distortions
  which originate from torques $\vec\mu\times\vec B$ of the Dy~momenta $\vec\mu$ induced by an external magnetic field $\vec B$.
  As the different local easy axes are not collinear, these torques emerge for any field direction.
  The resulting lattice distortions can be visualized by magnetostriction measurements.
  Figs.~\ref{kvb_dyto}(e) and (f) show the field-induced length changes of \dto for \bnne and \beee, respectively.
  In both cases, the \dto crystals elongate parallel to the magnetic-field direction (red squares), where the effect is more
  pronounced for \bnne.
  In the high-field region, \ie~for magnetic fields above the (almost) saturation of $M(B)$, this elongation is essentially
  linear in $B$ and temperature independent in the temperature range between 0.25 and \mbox{2 K} (not shown).
  Thus, the high-field magnetostriction cannot be attributed to the spin-ice system in \dto.
  For \beee [Fig.~\ref{kvb_dyto}(f)], we also measured the field-induced length changes perpendicular to $\vec B$,
  namely parallel to $[1\bar10]$ and $[11\bar2]$, and we find that the crystal contracts in both perpendicular directions, 
  but to a much smaller extent as compared to the elongation parallel to \beee .
  Thus, the magnetic field causes a strong distortion of the lattice due to torques of the large Dy~momenta which tend to align parallel to the external field.
  These lattice distortions influence the phononic properties,
  in particular, the phononic thermal conductivity.
  The steplike anomaly observed in $\kappa(B)$ for \bnne around \mbox{2 T} [Figs.~\ref{kvb_dyto}(a) and (b)] has,  indeed,
  an equivalent within the magnetostriction data, which show a small anomaly at the same magnetic field. 
  In Fig.~\ref{kvb_dyto}(e), an arrow marks this anomaly, which is better seen in the field-derivative of the length change [inset of Fig.~\ref{kvb_dyto}(e)]. The origin of this anomaly is, however, unclear.
  In sum, the field dependence of the thermal conductivity of \dto observed in the high-field region
  [\mbox{Figs. \ref{kvb_dyto}(a)-(d)}] can be explained
  by lattice distortions and can, thus, be attributed to a field-dependent phononic contribution \kappaphb.
  In addition, the weak, essentially linear high-field increase of the magnetization 
  observed above about 1.5~T [\cf~Fig.~\ref{kvb_b001}(a)] can also be explained by the torques $\vec\mu\times\vec B$. 
  For example, the additional increase of $M(B)$ between 2 and 7~T can be traced back to a tilting of the Dy~momenta by \mbox{$\sim1.5^\circ$}.

\subsection{Magnetic Heat Conductivity $\boldsymbol\kappa_\text{mag}$}  

\begin{figure}
    \includegraphics[width=\linewidth]{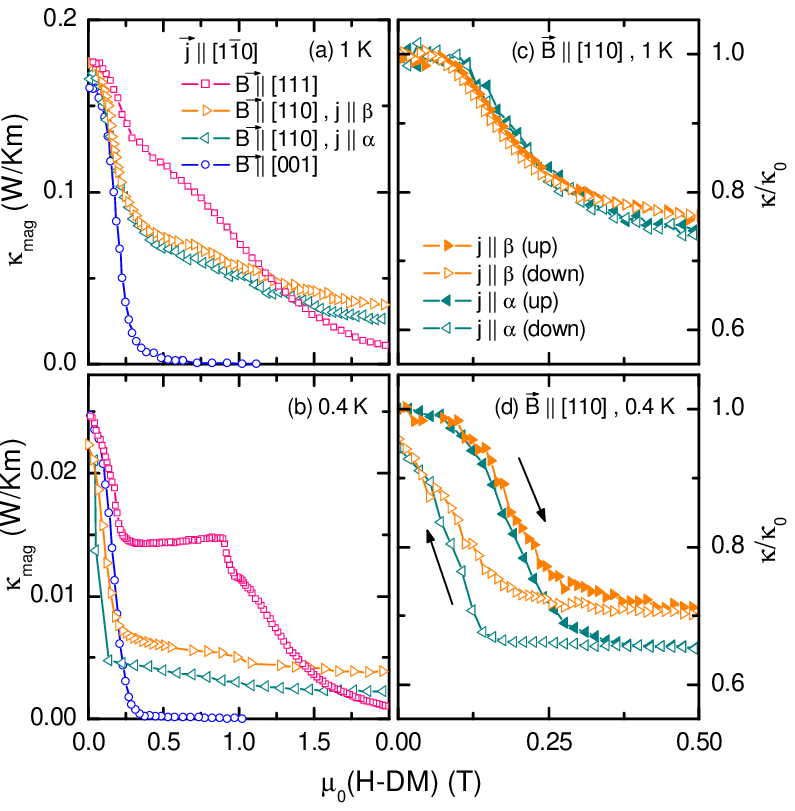}
    \caption{\label{kappamag}(Color online) (a) and (b): Field-dependence of \kappamag of \dto for different 
    magnetic-field directions at \mbox{1 K} and \mbox{0.4 K}, respectively.
    For clarity, only the data measured with decreasing field are shown.
    (c) and (d): Field-induced relative change \kapparellong for \been and different directions of the heat current $\vec j$
    at \mbox{1 K} and \mbox{0.4 K}, respectively.
    The arrows indicate the different field-sweep directions.
    Demagnetization effects are taken into account.}
  \end{figure}

  Now we proceed with the discussion of the magnetic contribution \kappamag of \dto.
  In the previous paragraphs, we showed that \kappamag can be suppressed by a rather small magnetic field,
  depending on the actual field direction.
  As \kappaph is only weakly field dependent for small magnetic fields,
  we can extract \kappamag by assuming a constant, \ie field-independent, phononic background [dashed line, \cf~Fig.~\ref{kvb_dyto}(b)].
  The field-dependent \kappamag for the different field directions are displayed in Figs.~\ref{kappamag}(a) and (b)
  at \mbox{1 K} and \mbox{0.4 K}, respectively.
  For clarity, we only show the data measured with decreasing magnetic field,
  which are (for $B>0$) closer to thermal equilibrium than the data obtained with increasing field (see discussion below).
  The magnetic contribution is maximum in zero field (\mbox{6-fold} degenerate ground state) and completely vanishes for \bnne (non-degenerate).
  For \been (\mbox{2-fold} degenerate), \kappamag cannot be completely suppressed, even for a rather large field of \simm{7}{T}.
  For this particular field direction, the crystallographic directions $[110]$ and $[1\bar10]$ become inequivalent.
  The spins with a non-vanishing component parallel to $\vec B$ become fully polarized and form so-called $\alpha$~chains
  running parallel to $\vec B$. The remaining spins are perpendicular to $\vec B$ and form $\beta$~chains that are oriented perpendicular to $\vec B$, \ie~$\alpha\,||\,[110]$ and $\beta\,||\,[1\bar10]$, \cf~Refs.~\onlinecite{Klemke2011, Yoshida2004, Fennell2002}.
  For \been, we measured \kappamag within two different measurement setups, either with the heat-current direction
  along the $\alpha$~chains, \ie $\vec j\,||\,\vec B$, or with the heat current along the $\beta$~chains, \ie $\vec j \perp \vec B$.
  At \mbox{1 K}, both datasets agree with each other [Fig.~\ref{kappamag}(a)],
  whereas at \mbox{0.4 K}, \kappamag is significantly larger along the $\beta$~chains [Fig.~\ref{kappamag}(b)].
  A direct comparison of the field-induced relative changes of $\kappa$ for $\vec j\,||\,\alpha$ and $\vec j\,||\,\beta$
  at \mbox{1 K} and \mbox{0.4 K} is displayed in Figs.~\ref{kappamag}(c) and (d), respectively.
  We note that the data of Figs.~\ref{kappamag}(c) and (d) have been obtained in a different measurement run
  on another crystal as compared to the data displayed in panels~(a) and (b). 
  The low-temperature (\mbox{0.4 K}) data for both heat-current directions show a pronounced hysteresis
  with respect to the field-sweep direction and the $\kappa(B\to 0)$ data measured with decreasing field end
  in a reduced zero-field $\kappa$~value, as it is also observed for \bnne and \beee.
  Moreover, the low-temperature data for \been clearly reveal that \kappamag is larger
  for a heat current along the $\beta$~chains than along the $\alpha$~chains.
  Such an anisotropy of \kappamag with respect to $\vec j$ can be qualitatively explained within the microscopic model
  of monopole excitations propagating via single spin flips.
  The spins of the $\alpha$~chains are easily polarized by \been, whereas the spins of the $\beta$~chains
  are perpendicular to $\vec B$ and are, thus, not affected.
  As a consequence, the monopole propagation along the $\alpha$~chains should be suppressed compared to the $\beta$~chains.
  Within the model of single spin flips, one would even expect that monopole excitations can only propagate along the $\beta$~chains,
  which are separated by (fully polarized) $\alpha$~chains, whereas \kappamag should completely vanish for $\vec j\,||\,\alpha$.
  A complete suppression of \kappamag along the $\alpha$~chains is, however, not observed [Figs.~\ref{kappamag}(a) and (b)].
  This shows that the single spin-flip formalism, which only accounts for nearest-neighbor interaction,
  is an oversimplification and the long-range dipolar interaction has to be taken into account.
  
  For a magnetic field parallel to $[111]$ within the \kagome-ice phase below \mbox{1 T} (3-fold degenerate),
  \kappamag is significantly larger than for \been [Figs.~\ref{kappamag}(a) and (b)].
  Above \mbox{1 T}, the ground-state degeneracy is lifted, resulting in a suppression of \kappamag,
  which essentially vanishes above \simm{1.5}{T} (indicated by a kink).
  As mentioned above, the field dependence of \kappaph cannot be neglected above \simm{1.5}{T}.
  Hence, the assumption of a field-independent phononic background is no longer justified,
  and \kappamag extracted here does not exactly approach zero.
  The hysteretic behavior below \mbox{1 T}, however, can certainly be attributed to \kappamag
  because no hysteresis of \kappaphb is observed for the reference compound \dytolong.
  At lowest temperature (below \simm{0.4}{K}),
  \kappamag within the \kagome-ice phase (below \mbox{1 T}) depends on whether the
  measurement is performed with increasing field after zero-field cooling or
  with decreasing field starting from the fully polarized phase (above \mbox{1 T}) (\cf~Fig.~\ref{kvb_b111}).
  The hysteresis in $\kappa(B)$ below \simm{0.3}{T} and the fact that the measurement with decreasing field does not end
  at the initial zero-field-cooled value is also present for \been [Fig.~\ref{kappamag}(d)] and
  for \bnne (\cf~Ref.~\onlinecite{Kolland2012}),
  and there are corresponding hysteresis effects in the respective magnetization data (inset of Fig.~\ref{kvb_b111}).
  
  The hysteresis within the \kagome-ice phase, however, has no equivalent within the magnetization data.
  The absence of such a hysteresis in $M(B)$ can be explained by the single-tetrahedron model.
  The 3-fold degenerate ground state consists of three different, energetically equivalent tetrahedron configurations.
  As these configurations have the same magnetization component parallel to \beee,
  they are indistinguishable for this particular field direction.
  The thermal conductivity, however, seems to be sensitive to the actually realized ground-state configuration.
  Above \mbox{1 T}, the spin ice is fully polarized, and the thermal conductivity does not depend on the
  actual field-sweep direction.
  When entering the \kagome-ice phase from higher field, the spins of the triangular planes
  which separate the \kagome planes stay polarized,
  whereas the spins of the \kagome planes arrange in such a way that the tetrahedra obey the ice rule \zz.
  As this configuration can easily be achieved by flipping only one spin per tetrahedron,
  it can be assumed that this configuration is close to thermal equilibrium.
  As a consequence, \kappamag stays almost constant on the \kagome-ice plateau.
  The situation is more complicated when approaching the \kagome-ice phase
  from below, \ie from the highly entropic ground state after zero-field cooling.
  The magnetic field \beee initially polarizes the spins of the triangular planes.
  Flipping such a spin, however, produces a monopole/anti-monopole pair in two neighboring \kagome planes.
  To reach thermal equilibrium, these monopoles (and anti-monopoles) have to be annihilated.
  This out-of-equilibrium state results in a reduced \kappamag, and time-dependent measurements of \kappamag (not shown) reveal that it very slowly converges towards the larger equilibrium value.
  When approaching the transition at \mbox{1 T}, the energy gap to the fully polarized state vanishes and, as a consequence, the \kagome planes can easily reach thermal equilibrium.
  This results in an increase of $\kappa(B)$ towards \mbox{1 T}.
  An interesting finding is the fact that in the field-decreasing run, \ie in (or at least close to) thermal equilibrium, 
  \kappamag remains constant on the \kagome-ice plateau although the energy gap to the (anti-)monopole excitations increases with decreasing field. 
    
  In sum, our data for the different magnetic-field directions reveal that there is a  magnetic contribution \kappamag,
  whose magnitude directly correlates with the degree of the ground-state degeneracy of the magnetic spin ice~\cite{footnote3}.
  At least qualitatively, we have evidence that \kappamag arises from the monopole excitations and that its magnitude mainly depends on the monopole mobility which is related to the degree of the ground-state degeneracy.
  However, on a more quantitative level there are many open questions. Hence, a more fundamental theory is required.

  \section{CONCLUSION}

  In conclusion, we find that the low-temperature thermal conductivity of the
  spin-ice compound \dto has a pronounced magnetic contribution \kappamag in zero magnetic field. 
  This \kappamag has a strong magnetic-field dependence, which is highly anisotropic with respect to the magnetic-field direction. 
  The magnitude of \kappamag directly correlates with the degree of the ground-state degeneracy.
  It is maximum in zero field, where the ground-state degeneracy is maximum (6-fold). Applying finite magnetic fields 
  along different directions stabilize other (spin-ice) ground states with reduced degrees of degeneracy depending on the direction and magnitude of the magnetic field. 
  A 3-fold degenerate \kagome-ice phase occurs for \beee below \mbox{1 T}, where \kappamag is reduced compared to the
  zero-field value, but remains larger than \kappamag in the 2-fold degenerate ground state which is realized for \been. 
  Non-degenerate ground states are reached either for  \bnne or for \beee above \mbox{1 T} and in both cases, \kappamag
  vanishes completely.
  For \been, we observe an additional anisotropy of \kappamag with respect to the direction of the heat current $\vec j$. 
  For a heat current directed along the $\alpha$~chains, whose spins are easily polarized because of their finite components along the field direction, \kappamag is suppressed more strongly as compared to a heat current $\vec j$ directed along the perpendicular oriented $\beta$~chains, whose spins are $\perp \vec B$ and are thus unaffected by the magnetic field. The anisotropy of \kappamag with respect to the direction of $\vec j$ is, however, too weak to be explained within   
  a single spin-flip formalism that only accounts for nearest-neighbor interaction. 
  
  Independent of the direction of the magnetic field,
  there are pronounced hysteresis effects when the magnetic field is reduced from larger values towards zero field in the low-temperature region.
  In all these cases, $\kappa(\vec B \rightarrow 0)$ approaches a reduced value compared to $\kappa_0$ obtained in the zero-field cooling process,
  and as a function of time, the reduced $\kappa(\vec B \rightarrow 0)$ value slowly relaxes towards $\kappa_0$.
  This low-field hysteresis of $\kappa(\vec B \rightarrow 0)$ correlates with the occurrence of a small,
  but finite remnant magnetization observed for all studied field directions at low temperature.        
  For \beee, however, we also observe a hysteresis of \kappamag that has no analogue in the magnetization data.
  Within the \kagome-ice phase, \kappamag strongly depends on whether this phase is entered from zero or from high magnetic fields. 
  Increasing the magnetic field from the degenerate zero-field ground state results
  in a \kappamag of the \kagome-ice phase that is strongly reduced compared to the value of \kappamag
  which is obtained when the \kagome-ice phase is entered from above, \ie from the fully polarized state starting from higher fields.
  This larger \kappamag is practically field-independent within the \kagome-ice phase and it neither depends on time,
  whereas the lower \kappamag slowly relaxes towards the larger plateau value.

  Furthermore, we also identify a considerable field dependence of the phononic contribution \kappaph which emerges at higher field,
  \ie~above the (almost) saturation of the magnetization.
  Essentially, the same field-dependent \kappaphb is observed in the half-doped reference compound \dytolong,
  where the spin-ice features are almost completely suppressed and all the above-described low-field features of \kappamag are essentially absent.   
  The field dependence of \kappaph can be explained by lattice distortions due to torques on the non-collinear Dy~momenta
  induced by the external magnetic field, and the resulting small tilts of the Dy~momenta can also explain the continuous increase of the high-field magnetization.   

\section{Acknowledgments}
    
  We acknowledge fruitful discussions with O.~Breunig, L.~Fritz, C.~Grams, J.~Hemberger, and A.~Rosch. This work has been   financially supported by the Deutsche Forschungsgemeinschaft.


\begin{thebibliography}{26}
\bibitem{footnote2}
  The magnetization has been measured with a continuous field-sweep rate of \mbox{$0.02 \,\text{T}/\text{min}$}.
  Because the thermal conductivity has been measured stepwise, the effective rate of the field change depends on the time needed to reach the stabilization criteria and typically amounts to \mbox{$\approx 0.01 \,\text{T}/\text{min}$}.
\bibitem{footnote3}
  The 3-fold degeneracy of the \kagome -ice phase can also be lifted by a weak misorientation of the magnetic field \beee ,   
  but for small angles the resulting energy splitting of the different 2in-2out states is small. Thus, we expect this effect to be of minor importance for the principle behavior of \kappamag in the plateau region, but it might be relevant for its unusual hysteresis behavior.
\end{thebibliography}
\end{document}